\documentstyle{article}
\begin{document}
\title{\bf Relativistic description of exclusive
heavy-to-light semileptonic decays $B\to\pi(\rho)e\nu$}
\author{ R. N. Faustov, V. O. Galkin and A. Yu. Mishurov\\
\small\it Russian Academy of Sciences, Scientific Council for
Cybernetics,\\
\small\it Vavilov Street 40, Moscow 117333, Russia}
\date{}
\maketitle

\begin{abstract}
The method of calculating  electroweak decay matrix elements between
heavy-heavy and heavy-light meson states is developed in the framework
of relativistic quark model based on the quasipotential approach in
quantum field theory. This method is applied for the study of
exclusive semileptonic $B\to\pi(\rho)$ decays. It is shown that the
large value of the final $\pi(\rho)$ meson recoil momentum allows for
the expansion in inverse powers of $b$-quark mass of the decay form
factors  at $q^2=0$, where $q^2$ is a momentum carried by the lepton
pair. This $1/m_b$ expansion considerably simplifies the analysis of
these decays and is carried out up to the second order. The
$q^2$-dependence of the form factors is investigated. It is found that
the $q^2$-behaviour of the axial form factor $A_1$ is different from
the other form factors. It is argued that the ratios $\Gamma(B\to\rho
e\nu)/\Gamma(B\to\pi e\nu)$ and $\Gamma_L/\Gamma_T$ are sensitive
probes of the $A_1$ $q^2$-dependence, and thus their
experimental measurement may discriminate between different
approaches.  We find $\Gamma(B\to\pi e\nu)=(3.0\pm0.6)
\times|V_{ub}|^2\times10^{12}$s$^{-1}$ and $\Gamma(B\to\rho e\nu)
=(5.4\pm 1.2)\times|V_{ub}|^2\times10^{12}$s$^{-1}$. The relation
between semileptonic and rare radiative $B$-decays is discussed.

\smallskip
\noindent PACS number(s): 12.39Ki, 13.20He
\end{abstract}

\section{INTRODUCTION}

The investigation of semileptonic decays of $B$ mesons into light
mesons is important for the determination of the
Cabibbo-Kobayashi-Maskawa matrix element $V_{ub}$, which is the most
poorly studied. At present the value of $V_{ub}$ is mainly determined
from the endpoint of the lepton spectrum in semileptonic $B$-decays
\cite{1}. Unfortunately, the theoretical interpretation of the
endpoint region of the lepton spectrum in inclusive $B\to X_u\ell \bar
\nu $ decays is very complicated and suffers from large uncertainties
\cite{2}.  The other way to determine $V_{ub}$ is to consider
exclusive semileptonic decays $B\to \pi(\rho)e\nu$. These are the
heavy-to-light transitions with a wide kinematic range. In contrast
to the heavy-to-heavy transitions, here we can not expand matrix
elements in the inverse powers of the final quark mass. It is also
necessary to mention that the final meson has a large recoil momentum
almost in the whole kinematical range. Thus the motion of final
$\pi(\rho)$ meson should be treated relativistically. If we consider
the point of maximum recoil of the final meson, we find that
$\pi(\rho)$ bears the large relativistic recoil momentum $\vert{\bf
\Delta}_{max}\vert$ of order  $m_b/2$ and the energy of the same
order. Thus at this kinematical point it is possible to expand the
matrix element of the weak current both in inverse powers of
$b$-quark mass of the initial $B$ meson and in inverse powers of
the recoil momentum $\vert {\bf \Delta}_{max}\vert$ of the final
$\pi(\rho)$ meson. As a result the expansion in powers $1/m_b$ arises
for the $B\to \pi(\rho)$ semileptonic form factors at $q^2=0$, where
$q^2$ is a momentum carried  by the lepton pair. The aim of this
paper is to realize such expansion in the framework of relativistic
quark model. We show that this expansion considerably simplifies the
analysis of exclusive $B\to \pi(\rho)e\nu$ semileptonic decays.

Our relativistic quark model is based on the quasipotential approach
in
quantum field theory with the specific choice of the $q\bar q$
potential.  It provides a consistent scheme for calculation of all
relativistic corrections at a given order of $v^2/c^2$ and allows for
the heavy quark $1/m_Q$ expansion. This model has been applied for the
calculations of meson mass spectra \cite{3}, radiative decay widths
\cite{4}, pseudoscalar decay constants \cite{5}, heavy-to-heavy
semileptonic \cite{6} and nonleptonic \cite{7} decay rates. The heavy
quark $1/m_Q$ expansion in our model for the heavy-to-heavy
semileptonic transitions has been developed in \cite{8} up to
$1/m_Q^2$ order. The results are in agreement with the model
independent predictions of the heavy quark effective theory (HQET)
\cite{9}. The $1/m_b$ expansion of rare radiative decay form factors
of $B$ mesons has been carried out in \cite{10} along the same lines
as in the present paper. We have briefly presented the results for
$B\to\pi e\nu$ and $B\to \rho e\nu$ decays in ref.~\cite{10a}, where
the expansion up to the first order in $1/m_b$ has been carried out.
In the present paper we extend the analysis up to the second order and
give a detailed discussion of the expansion method and results.

The  paper is organized as follows. The relativistic quark model is
described in Sect.~2. In Sect.~3 we give the detailed description of
the method of calculating decay matrix elements between heavy-heavy
and heavy-light meson states, based on the quasipotential approach. We
show that the heavy-to-heavy decay matrix elements can be expanded in
inverse powers of the heavy quark masses at  zero
recoil of the final meson. On the other hand, the heavy-to-light decay
matrix elements can be expanded in inverse powers of the initial heavy
quark mass at the maximum recoil of
the final light meson. These expansions permit the calculation of
decay matrix elements with the account of relativistic effects. In
Sect.~4 the method is applied to the calculation of the semileptonic
$B\to\pi(\rho)e\nu$ decay form factors. Our numerical
results for the form factors and decay rates are presented in
Sect.~5. Therein we  discuss the $q^2$-dependence of the form factors
and the relations between semileptonic $B\to\rho e\nu$  and rare
radiative $B\to\rho(K^*)\gamma$ decays. Sect.~6 contains our
conclusions. The formulae for the form factors at the point of maximum
recoil of the final light meson are given in Appendix.

\section{RELATIVISTIC QUARK MODEL}
In the quasipotential approach \cite{11} meson with the mass $M$ and
relative momentum of quarks {\bf p} is described by
a single-time quasipotential wave function $\Psi_M({\bf p})$,
projected onto positive-energy states. This wave function satisfies
the quasipotential equation
\begin{equation} \label{1} \left(M-({\bf
p}^2+m_a^2)^{1/2}-({\bf p}^2+m_b^2)^{1/2}\right) \Psi_M({\bf p}) =
\int \frac{d^3 q}{(2\pi)^3} V({\bf p},{\bf q};M)\Psi_M({\bf q}),
\end{equation}
The quasipotential equation (\ref{1}) can be
transformed into a local Schr\"odinger-like equation~\cite{12}
\begin{equation} \label{2} \Big(\frac{b^2(M)}{2\mu_{R}}-\frac{{\bf
p}^2}{2\mu_{R}}\Big)\Psi_{M}({\bf p})=\int\frac{d^3 q}{(2\pi)^3}
V({\bf p,q};M)\Psi_{M}({\bf q}),\end{equation}
where the relativistic reduced mass is
\begin{equation} \label{3}
\mu_{R}=\frac{M^4-(m^2_a-m^2_b)^2}{4M^3};\end{equation}
and the square of the relative momentum on the mass shell is
\begin{equation} \label{4}
b^2(M)=\frac{[M^2-(m_a+m_b)^2][M^2-(m_a-m_b)^2]}{4M^2}, \end{equation}
$m_{a,b}$ are the quark masses. While constructing the kernel of this
equation $V({\bf p,q};M)$ --- the quasipotential of quark-antiquark
interaction --- we have assumed that effective interaction is the sum
of the one-gluon exchange term with the mixture of long-range vector
and scalar linear confining potentials. We have also assumed that at
large distances quarks acquire universal nonperturbative anomalous
chromomagnetic moments and thus the vector long-range potential
contains the Pauli interaction. The quasipotential is defined by
\cite{3}:
\begin{equation} \label{5} V({\bf p,q},M)=\bar{u}_a(p)
\bar{u}_b(-p)\Big\{\frac{4}{3}\alpha_SD_{ \mu\nu}({\bf
k})\gamma_a^{\mu}\gamma_b^{\nu}+V^V_{conf}({\bf k})\Gamma_a^{\mu}
\Gamma_{b;\mu}+V^S_{conf}({\bf
k})\Big\}u_a(q)u_b(-q),\end{equation}
where $\alpha_S$ is the QCD coupling constant, $D_{\mu\nu}$ is the
gluon propagator; $\gamma_{\mu}$ and $u(p)$ are the Dirac matrices and
spinors; ${\bf k=p-q}$; the effective long-range vector vertex is
\begin{equation} \label{6} \Gamma_{\mu}({\bf k})=\gamma_{\mu}+
\frac{i\kappa}{2m}\sigma_{\mu\nu}k^{\nu},\end{equation}
$\kappa$ is the anomalous chromomagnetic quark moment. Vector and
scalar confining potentials in the nonrelativistic limit reduce to
\begin{equation} \label{7} V^V_{conf}(r)=(1-\varepsilon)(Ar+B),\ \
V^S_{conf}(r)=\varepsilon(Ar+B),\end{equation}
reproducing $V_{nonrel}^{conf}(r)=V^S_{conf}+V^V_{conf}=Ar+B$, where
$\varepsilon$ is the mixing coefficient. The explicit expression for
the quasipotential with the account of the relativistic corrections of
order $v^2/c^2$ can be found in ref. \cite{3}.  All the parameters of
our model: quark masses, parameters of linear confining potential $A$
and $B$, mixing coefficient $\varepsilon$ and anomalous chromomagnetic
quark moment $\kappa$ were fixed from the analysis of meson masses
\cite{3} and radiative decays \cite{4}.  Quark masses: $m_b=4.88$ GeV;
$m_c=1.55$ GeV; $m_s=0.50$ GeV; $m_{u,d}=0.33$ GeV and parameters of
linear potential: $A=0.18$ GeV$^2$; $B=-0.30$ GeV have standard values
for quark models.  The value of the mixing coefficient of vector and
scalar confining potentials $\varepsilon=-0.9$ has been primarily
chosen from the consideration of meson radiative decays, which are
very sensitive to the Lorentz-structure of the confining potential:
the resulting leading relativistic corrections coming from vector and
scalar potentials have opposite signs for the radiative Ml-decays
\cite{4}. Universal anomalous chromomagnetic moment of quark
$\kappa=-1$ has been fixed from the analysis of the fine splitting of
heavy quarkonia ${ }^3P_J$- states \cite{3}.

Recently we have considered the expansion of the matrix elements of
weak
heavy quark currents between pseudoscalar and vector meson states up
to
the second order in inverse powers of the heavy quark masses \cite{8}.
It has been found that the general structure of leading, subleading
and second order $1/m_Q$ corrections in our relativistic model is in
accord with the predictions of HQET. The heavy quark symmetry and QCD
impose rigid constraints on the parameters of the long-range potential
of our model. The analysis of the first order corrections \cite{8}
allowed to fix the value of effective long-range anomalous
chromomagnetic moment of quarks $ \kappa =-1$, which coincides with
the result, obtained from the mass spectra \cite{3}. The mixing
parameter of vector and scalar confining potentials has been found
from the comparison of the second order corrections to be $
\varepsilon =-1$. This value is very close to the previous one
$\varepsilon =-0.9$ determined from radiative decays of mesons
\cite{4}. Therefore, we have got QCD and heavy quark symmetry
motivation for the choice of the main parameters of our model. The
found values of $\varepsilon$ and $\kappa$ imply that confining
quark-antiquark potential has predominantly Lorentz-vector structure,
while the scalar potential is anticonfining and helps to reproduce the
initial nonrelativistic potential.

\section{MATRIX ELEMENTS OF ELECTROWEAK CURRENT BETWEEN HEAVY-HEAVY
AND HEAVY-LIGHT MESON STATES}
The matrix element of the local current $J$ between bound states in
the quasipotential method has the form \cite{13}
\begin{equation}\label{8} \langle M' \vert J_\mu (0) \vert M\rangle
=\int \frac{d^3p\, d^3q}{(2\pi )^6} \bar \Psi_{M'}({\bf
p})\Gamma _\mu ({\bf p},{\bf q})\Psi_M({\bf q}),\end{equation}
where $M(M')$ is initial (final) meson, $\Gamma _\mu ({\bf p},{\bf
q})$ is the two-particle vertex function and  $\Psi_{M,M'}$ are the
meson wave functions projected onto the positive energy states of
quarks.

This relation is valid for the general structure of the current
$J=\bar Q'GQ$, where $G$ can be an arbitrary combination of Dirac
matrices. The contributions to $\Gamma$ come from Figs.~1 and 2.  Thus
the vertex functions look like
\begin{equation} \label{9}\Gamma^{(1)}({\bf
p},{\bf q})=\bar u_{Q'}(p_1)G u_Q(q_1)(2\pi)^3\delta({\bf p}_2-{\bf
q}_2),\end{equation}
and
\begin{eqnarray}\label{10} \Gamma^{(2)}({\bf
p},{\bf q})&=&\bar u_{Q'}(p_1)\bar u_q(p_2) \Bigl\{G
\frac{\Lambda_Q^{(-)}(
k_1)}{\varepsilon_Q(k_1)+\varepsilon_Q(p_1)}\gamma_1^0V({\bf p}_2-{\bf
q}_2)\nonumber \\ & &+V({\bf p}_2-{\bf
q}_2)\frac{\Lambda_{Q'}^{(-)}(k_1')}{ \varepsilon_{Q'}(k_1')+
\varepsilon_{Q'}(q_1)}\gamma_1^0 G\Bigr\}u_Q(q_1)
u_q(q_2),\end{eqnarray}
where ${\bf k}_1={\bf p}_1-{\bf\Delta};\qquad
{\bf k}_1'={\bf q}_1+{\bf\Delta};\qquad {\bf\Delta}={\bf p}_M-{\bf
p}_{M'}; \qquad \varepsilon (p)=(m^2+{\bf p}^2)^{1/2}$;
$$\Lambda^{(-)}(p)=\frac{\varepsilon(p)-\bigl( m\gamma
^0+\gamma^0({\bf \gamma p})\bigr)}{ 2\varepsilon (p)}.$$
and
\begin{eqnarray*} p_{1,2}&=&\varepsilon_{1,2}(p)\frac{p_{M'}}{M'}
\pm\sum_{i=1}^3 n^{(i)}(p_{M'})p^i,\\
q_{1,2}&=&\varepsilon_{1,2}(p)\frac{p_M}{M} \pm \sum_{i=1}^3 n^{(i)}
(p_M)q^i,\end{eqnarray*}
here
$$ n^{(i)\mu}(p)=L_{p_i}^\mu=\left\{ \frac{p^i}{M},\ \delta_{ij}+
\frac{p^ip^j}{M(E+M)}\right\},$$
Note that the contribution $\Gamma^{(2)}$ is the consequence
of the projection onto the positive-energy states. The form of the
relativistic corrections resulting from the vertex function
$\Gamma^{(2)}$ is explicitly dependent on the Lorentz-structure of
$q\bar q$-interaction.

The general structure of the current matrix element (\ref{8}) is
rather complicated, because it is necessary to integrate both with
respect to $d^3p$ and $d^3q$. The $\delta$-function in the expression
(\ref{9}) for the vertex function $\Gamma^{(1)}$ permits to perform
one of these integrations. As a result the contribution of
$\Gamma^{(1)}$ to the current matrix element has usual structure and
can be calculated without any expansion, if the wave functions of
initial and final meson are known. The situation with the contribution
$\Gamma^{(2)}$ is different. Here instead of $\delta$-function we have
a complicated structure, containing the potential of $q\bar
q$-interaction in meson. Thus in general case we cannot get rid of one
of the integrations in the contribution of $\Gamma^{(2)}$ to the
matrix element (\ref{8}). Therefore, it is necessary to use some
additional considerations. The main idea is to expand the vertex
function $\Gamma^{(2)}$, given by (\ref{10}), in such  a way that it
will be possible to use the quasipotential equation (\ref{1}) in order
to perform one of the integrations in the current matrix element
(\ref{8}). The realization of such expansion  differs for the cases of
heavy-to-heavy and heavy-to-light transitions.

\subsection{Heavy-to-heavy decay matrix elements}
At first we consider the heavy-to-heavy meson decays, such as
semileptonic $B\to De\nu$ decays, and radiative transitions in
quarkonia and $B^*\to B\gamma$, $D^*\to D\gamma$. Here we have two
natural expansion parameters, which are the heavy quark masses in
initial and final meson. The most convenient point for the expansion
of vertex function $\Gamma^{(2)}$ in inverse powers of the heavy quark
masses for semileptonic decays is the point of zero recoil of final
meson, where ${\bf\Delta}=0$. For radiative decays the momentum
transfer is fixed $\vert{\bf\Delta}\vert=\frac{M_M^2-M_{M'}^2}{2M}$.
The difference of initial and final meson masses is proportional to
the fine or hyperfine splitting and thus $\vert{\bf\Delta}\vert/M =
o(1/M^2)$, so zero recoil is a good approximation.

It is easy to see that $\Gamma^{(2)}$ contributes to the current
matrix element at first order of $1/m_Q$ expansion for transitons
between mesons consisting from heavy and light quarks ($B$, $D$
mesons) \cite{8} and at second order of $v/c$ expansion for mesons
consisting from two heavy quarks of the same flavour (quarkonia
$\Upsilon$, $J/\Psi$) \cite{4}. We limit our analysis to the
consideration of the terms up to the second order in $1/m_Q$ or $v/c$
expansions.  We substitute the Dirac matrices $G$ and spinors $u$ in
the vertex function $\Gamma^{(2)}$ and consider the cases of
Lorentz-scalar and Lorentz-vector (with Pauli term) $q\bar
q$-interaction potential. Then we expand $\Gamma^{(2)}$ to the desired
order and see that it is possible to integrate either with respect to
$d^3p$ or $d^3q$ in the current matrix element (\ref{8}) using
quasipotential equation (\ref{1}). Performing these integrations and
taking the sum of the contributions of $\Gamma^{(1)}$ and
$\Gamma^{(2)}$ we get the expression for the current matrix element,
which contains the ordinary mean values between meson wave functions.
Thus this matrix element can be easily calculated numerically if the
meson wave functions are known. The described method  has been applied
to the calculations of heavy-to-heavy semileptonic decays in
\cite{6,8} and radiative decays in \cite{4}.

\subsection{Heavy-to-light decay matrix elements}
Now we consider the heavy-to-light meson decays, such as semileptonic
$B\to \pi(\rho)e\nu$ and rare radiative $B\to K^*\gamma$ decays. In
these decays the final meson contains only light quarks ($u$, $d$,
$s$), thus, in contrast to the heavy-to-heavy transitions, we cannot
expand matrix elements in inverse powers of the final quark mass. The
expansion of $\Gamma^{(2)}$ only in inverse powers of the initial
heavy quark mass at ${\bf\Delta}=0$ does not allow to use the
quasipotential equation for performing one of the integrations in
corresponding current matrix element (\ref{8}). However, as it was
already mentioned in the introduction, the final light meson has the
large recoil momentum almost in the whole  kinematical range. At the
point of maximum recoil of final light meson \footnote{In the case of
rare radiative decays the recoil momentum of final light meson is
fixed at the maximum value ${\bf\Delta}_{max}$.} the large value of
recoil momentum ${\bf\Delta}_{max}\sim m_Q/2$ allows for the
expansion of decay matrix element in $1/m_Q$. The contributions to
this expansion come both from the inverse powers of heavy $m_Q$ from
initial meson and from inverse powers of the recoil momentum
$|{\bf\Delta}_{max}|$ of the final light meson. The large value of
recoil momentum $|{\bf\Delta}_{max}|$ permits to neglect ${\bf p}^2$
in comparison with ${\bf\Delta}_{max}^2$ in the light quark energy
$\varepsilon_{q,Q'}(p+\Delta)$ in final meson in the expression for
the matrix element originating from $\Gamma^{(2)}$. Such approximation
corresponds to omitting terms of the third order in $1/m_Q$
expansion and is compatible with our analysis, which is carried out up
to the second order. It is easy to see that we can now perform one of
the integrations in the current matrix element (\ref{8}) using the
quasipotential equation as in the case of heavy final meson. As a
result we again get the expression for the current matrix element,
which contains only the ordinary mean values between meson wave
functions, but in this case at the point of maximum recoil of final
light meson. This method has been applied to calculation of rare
radiative decays of $B$ mesons in ref.~\cite{10} and in the next
section we use it for consideration of $B\to\pi(\rho)e\nu$
semileptonic decays.

\section{$B\to \pi(\rho)e\nu $ DECAY FORM FACTORS}

\subsection{Decay form factors at $q^2=0$}
The form factors of the semileptonic decays $B\to \pi e\nu$ and $B\to
\rho e\nu$ are defined in the standard way as:
\begin{eqnarray}\label{11}\langle\pi(p_\pi)\vert \bar q\gamma_\mu
b\vert B(p_B) \rangle& =& f_+(q^2)(p_B+p_\pi)_\mu +
f_-(q^2)(p_B-p_\pi)_\mu,\\
\label{12}\langle\rho(p_\rho,e)\vert \bar
q\gamma_\mu(1-\gamma^5) b \vert B(p_B)\rangle &=&
-(M_B+M_\rho)A_1(q^2)e^*_\mu + \frac{A_2(q^2)}{M_B+M_\rho}
(e^* p_B)(p_B+p_\rho)_\mu\nonumber\qquad\qquad\hfil\hfill\\ & &
+\frac{A_3(q^2)}{M_B+M_\rho}(e^*
p_B)(p_B-p_\rho)_\mu + \frac{2V(q^2)}{M_B+M_\rho}i\epsilon_{\mu\nu
\tau\sigma}{e^*}^\nu p_B^\tau p_\rho^\sigma,\end{eqnarray}
where $q=p_B-p_{\pi(\rho)}$, $e$ is a polarization vector of $\rho $
meson. In the limit of vanishing lepton mass, the form factors $f_-$
and $A_3$ do not contribute to the decay rates and thus will not be
considered.

It is convenient to consider the decay $B\to\pi(\rho)e\nu$ in the $B$
meson rest frame. Then the wave function of the final $\pi(\rho)$
meson
moving with the recoil momentum ${\bf\Delta}$ is connected with the
wave function at rest by the transformation \cite{13}
\begin{equation}\label{13}\Psi_{\pi(\rho)\,{\bf\Delta}}({\bf
p})=D_q^{1/2}(R_{L{\bf\Delta}}^W)D_q^{1/2}(R_{L{
\bf\Delta}}^W)\Psi_{\pi(\rho)\,{\bf 0}}({\bf p}),\end{equation}
where $D^{1/2}(R)$ is the well-known rotation matrix and $R^W$ is the
Wigner rotation.

The meson wave functions in the rest frame have been calculated by
numerical
solution of the quasipotential equation (\ref{2}) \cite{14}. However,
it is more convenient to use analytical expressions for meson wave
functions. The examination of numerical results for the ground state
wave functions of mesons containing at least one light quark has shown
that they can be well approximated by the Gaussian functions
\begin{equation}\label{14}\Psi_M({\bf p})\equiv \Psi_{M\,{\bf 0}}({\bf
p})=\Bigl({4\pi\over \beta_M^2} \Bigr)^{3/4}\exp\Bigl(-{{\bf p}^2\over
2\beta_M^2}\Bigr),\end{equation}
with the deviation less than 5\%.

The parameters are

$$\beta_B=0.41\ {\rm GeV};\qquad\beta_{\pi(\rho)}=0.31\ {\rm GeV}.$$

Now we apply the method for calculation of decay matrix elements,
described in the previous section. At the point of maximal recoil
of final light meson
\begin{equation}\label{15} |{\bf \Delta}_{max}| = \frac{M_B^2-
M_{\pi(\rho)}^2}{2M_B}, \end{equation}
we expand the vertex function $\Gamma^{(2)}$ for the Lorentz-scalar
and Lorentz-vector (with Pauli term) $q\bar q$-interactions up to the
second order in $1/m_b$. Then we substitute the vertex functions
$\Gamma^{(1)}$ and $\Gamma^{(2)}$ in the matrix element (\ref{8}) and
take into account the Lorentz transformation of the final meson wave
function (\ref{13}). Performing one of the integrations in the current
matrix element (\ref{8}) (using the $\delta$-function in
$\Gamma^{(1)}$ and the quasipotential equation in the contribution of
$\Gamma^{(2)}$) we get for the form factors of $B\to\pi e\nu$ and
$B\to\rho e\nu$ decays the following expressions at $q^2=0$ point
\begin{eqnarray}
\label{16}
f_+(0)&=&f_+^{(1)}(0)+\varepsilon f_{+}^{S(2)}(0)+(1-\varepsilon)
f_{+}^{V(2)}(0),\\
\label{17}
A_1(0)&=&A_1^{(1)}(0)+\varepsilon A_{1}^{S(2)}(0)+(1-\varepsilon)
A_{1}^{V(2)}(0),\\
\label{18}
A_2(0)&=&A_2^{(1)}(0)+\varepsilon A_{2}^{S(2)}(0)+(1-\varepsilon)
A_{2}^{V(2)}(0),\\
\label{19}
V(0)&=&V^{(1)}(0)+\varepsilon V^{S(2)}(0)+(1-\varepsilon)
V^{V(2)}(0),
\end{eqnarray}
where $f_+^{(1)}$, $f_{+}^{S,V(2)}$, $A_{1,2}^{(1)}$,
$A_{1,2}^{S,V(2)}$, $V^{(1)}$ and $V^{S,V(2)}$ are given in Appendix,
the superscripts ``(1)" and ``(2)" correspond to Figs.~1 and 2, S and
V --- to the scalar and vector potentials of $q\bar q$-interaction.

\subsection{$1/m_b$ expansion for decay form factors}
Let us proceed further and for the sake of consistensy carry out the
complete expansion of form factors (\ref{16})--(\ref{19}) in inverse
powers of $b$-quark mass. For this expansion we will use some model
independent results obtained in HQET \cite{9}.

In HQET the mass of $B$ meson has the following expansion in $1/m_b$
\cite{9}
\begin{equation}\label{20}M_B=m_b+\bar\Lambda+\frac{\Delta m_B^2}{
2m_b}+O\left(\frac{1}{ m_b^2}\right),\end{equation}
where parameter $\bar\Lambda$ is the difference between the meson and
quark masses in the limit of infinitely heavy quark mass. In our model
$\bar\Lambda $ is equal to the mean value of light quark energy inside
the heavy meson $\bar\Lambda=\langle\varepsilon_q\rangle_B\approx
0.54$~GeV \cite{8}. $\Delta m_B^2$ arises from the first-order power
corrections to the HQET Lagrangian and has the form \cite{9}:
\begin{equation}\label{21}\Delta
m_B^2=-\lambda_1-3\lambda_2.\end{equation}
The parameter $\lambda_1$ results from the mass shift due to the
kinetic operator, while $\lambda_2$ parameterizes the chromomagnetic
interaction \cite{9}. The value of spin-symmetry breaking parameter
$\lambda_2$ is related to the vector-pseudoscalar mass splitting
$$\lambda_2\approx {1\over 4}(M_{B^*}^2-M_B^2)=0.12\pm 0.01\ {\rm
GeV}^2.$$
The parameter $\lambda_1$ is not directly connected with
observable quantities. Theoretical predictions for it vary in a wide
range:  $\lambda_1=-0.30\pm 0.30 \ {\rm GeV}^2$ \cite{9,15}.

In the limit $m_Q\to \infty$, meson wave functions become independent
of the flavour of heavy quark. Thus the Gaussian parameter $\beta_B$
in (\ref{14}) should have the following expansion \cite{8}
\begin{equation}\label{22}\beta_B=\beta-\frac{\Delta\beta^2}{
m_b}+O\left(\frac{1}{ m_b^2}\right),\qquad \beta \approx  0.42 \
{\rm GeV},\end{equation}
where the second term breaks the flavour symmetry and in our model is
equal to $\Delta\beta^2\approx 0.045\ {\rm GeV}^2$ \cite{8}.

Substituting (\ref{20}) in (\ref{15}) and (\ref{a13}) we get the
$1/m_b$ expansion of the recoil momentum and the energy of final
vector meson:
\begin{eqnarray}\label{23}\vert{\bf\Delta}_{max}\vert&=&\frac{m_b}{
2}\left(1+\frac{1}{m_b}\bar\Lambda +\frac{1}{ m_b^2}\left(\frac{\Delta
m_B^2}{ 2}-M_{\pi(\rho)}^2\right)\right)+O\left(\frac{1}{m_b^2}\right)
,\nonumber\\
E_{\pi(\rho)}&=&\frac{m_b}{ 2}\left(1+\frac{1}{ m_b}\bar
\Lambda+\frac{1}{ m_b^2}\left(\frac{\Delta m_B^2}{2}
+M_{\pi(\rho)}^2\right)\right)+O\left(\frac{1}{
m_b^2}\right).\end{eqnarray}

Now we use the Gaussian approximation for the wave functions
(\ref{14}).  Then shifting the integration variable ${\bf p}$ in
(\ref{a1})--(\ref{a12}) by $-\frac{\varepsilon_q}{
E_{\pi(\rho)}+M_{\pi(\rho})}{\bf\Delta}_{max}$, we can factor out the
${\bf\Delta}_{max}$ dependence of the meson wave function overlap in
form factors $f_+$, $A_{1,2}$, $V$. The result can be written in the
form
\begin{equation}\label{24}f_+(0)={\cal
F}_+({\bf\Delta}_{max}^2)\exp(-\zeta {\bf
\Delta}_{max}^2),\end{equation}
\begin{equation}\label{25}A_{1,2}(0)={\cal
A}_{1,2}({\bf\Delta}_{max}^2)\exp(-\zeta {\bf
\Delta}_{max}^2),\end{equation}
\begin{equation}\label{26}V(0)={\cal
V}({\bf\Delta}_{max}^2)\exp(-\zeta {\bf
\Delta}_{max}^2),\end{equation}
where $\vert{\bf\Delta}_{max}\vert$ is  given by (\ref{15}) and
\begin{equation}\label{27}\zeta{\bf\Delta}_{max}^2=
\frac{2\tilde\Lambda^2
{\bf\Delta}_{max}^2}{(\beta_B^2+\beta_{\pi(\rho)}^2)(E_{\pi(\rho)}+
M_{\pi(\rho)})^2} =\frac{\tilde\Lambda^2}{ \beta_B^2
}\eta\left(\frac{M_B-M_{\pi(\rho)}}{M_B+M_{\pi(\rho)}}\right)^2,
\end{equation}
here $\eta = \frac{2\beta_B^2}{\beta_B^2+\beta_{\pi(\rho)}^2}$ and
$\tilde\Lambda$ is equal to the mean value of light quark energy
between $B$ and $\pi(\rho)$ meson states:
\begin{equation}\label{28}\tilde\Lambda =\langle\varepsilon_q\rangle
\approx 0.53\ {\rm GeV}, \end{equation}

Expanding (\ref{27}) in powers of $1/m_b$ we get
\begin{equation}\label{29}\zeta{\bf\Delta}_{max}^2
=\frac{\tilde\Lambda^2}{ \beta^2}\eta\left(1-4{M_{\pi(\rho)}\over
m_b}\right)+O\left(\frac{1}{ m_b^2}\right).  \end{equation}
We see that the first term in this expansion for the decay into $\rho$
meson is large. Really,  $4{M_\rho / m_b}\approx 0.63$.  The value of
this correction is also increased by the exponentiating in
(\ref{24})--(\ref{26}). Therefore, we conclude that the first order
correction in $1/m_b$ expansion for the form factors of
$B\to\pi(\rho)e\nu$ decay, arising from the meson wave function
overlap, is large.\footnote{The same situation occurs for the rare
radiative $B$ decays \cite{10}.} Thus, taking into account that our
method of calculating decay matrix elements does not require the
expansion of the meson wave function overlap, we use unexpanded
expression (\ref{27}) in the exponential of the form factors
(\ref{24})--(\ref{26}).

In contrast to the meson wave function overlap the factors ${\cal
F}_1({\bf \Delta}_{max}^2) $, ${\cal A}_{1,2}({\bf\Delta}_{max}^2)$
and ${\cal V}({\bf\Delta}_{max}^2)$, which originate from the vertex
functions $\Gamma^{(1),(2)}$ and Lorentz-transformation
(\ref{13}) of the final meson wave function, have a well defined
$1/m_b$ expansion. The first and second order corrections are small.
Substituting the Gaussian wave functions (\ref{13}) in the expressions
for the form factor (\ref{16})--(\ref{19}) and
(\ref{a1})--(\ref{a12}), with the value of anomalous chromomagnetic
quark moment $\kappa=-1$, and using (\ref{24})--(\ref{26}) and the
expansions (\ref{20})--(\ref{23}), we get up to the second order in
$1/m_b$ expansion:
\nopagebreak \\ \medskip  \nopagebreak
\noindent a) $B\to\pi e\nu$ decay
\nopagebreak
\begin{eqnarray}\label{30}{\cal F}_+({\bf\Delta}_{max}^2)&=&{\cal
F}_+^{(1)}({\bf\Delta}_{max}^2)+ \varepsilon{\cal
F}_+^{S(2)}({\bf\Delta}_{max}^2)+(1-\varepsilon){\cal F}_+^{V(2)}({\bf
\Delta}_{max}^2);\\
\label{31}
{\cal F}_1^{(1)}({\bf\Delta}_{max}^2)&=&N\biggl\{1+\frac{1}{
m_b}X_1+\frac{1}{m_b^2}\biggl(Y_+-2\langle{\bf p}^2\rangle
-\frac{3}{4}\tilde\Lambda^2\eta^2+\tilde\Lambda\eta(2m_q+M_\pi)
\nonumber\\
& & +\frac{2}{3}\left\langle\frac{{\bf p}^2}{\bar\varepsilon_q+m_q}
\right\rangle \left(2\tilde\Lambda\eta-m_q-2M_\pi-
\frac{1}{2}\bar\Lambda\right) \biggr)\biggr\};\\
\label{32}
{\cal F}_+^{S(2)}({\bf\Delta}_{max}^2) &=&N\frac{1}{m_b^2}\biggl\{
-2m_q(M_\pi-2\langle\bar\varepsilon_q\rangle) -\frac{1}{3}\left\langle
\frac{{\bf p}^2}{\bar\varepsilon_q+m_q}\right\rangle(M_\pi(1-R)
+2m_q(1-2R)-2\bar\Lambda R)\nonumber\\
& &+\frac{2}{3}\langle{\bf p}^2\rangle(1-2R)+\frac{1}{2}\tilde\Lambda
\eta\left(2\langle\bar\varepsilon_q\rangle(1-2R)-M_\pi(1-R)
+2\bar\Lambda R  -\frac{1}{3}\left\langle\frac{{\bf p}^2}{\bar
\varepsilon_q}\right\rangle\right)\biggr\};\\
\label{33}
{\cal F}_+^{V(2)}({\bf\Delta}_{max}^2) &=&N \biggl\{-\frac{1}{m_b}
(2RZ_1+(1+2R)Z_2) +\frac{1}{m_b^2}\biggl( -\frac{2}{3}m_qRZ_3
+4\langle{\bf p}^2\rangle\left(1-\frac{3}{4}R\right)\nonumber\\
& & -\left\langle\frac{{\bf p}^2}{\bar\varepsilon_q+m_q}\right\rangle
\left( 2M_\pi
\left(1-\frac{1}{3}R\right) +4m_q\left(1-\frac{5}{4}R\right)
+\frac{5}{3}\bar\Lambda R\right)\nonumber\\
& & +\left(\left(2M_\pi
+\eta\frac{\beta_\pi^2}{\beta^2}\frac{\Delta\beta^2}{\beta}\right)
(1+2R)-2m_q(1-2R)\right)Z_2\biggr)\biggr\};
\end{eqnarray}
\noindent b) $B\to\rho e\nu$ decay
\nopagebreak
\begin{eqnarray}\label{34}{\cal A}_1({\bf\Delta}_{max}^2)&=&{\cal
A}_1^{(1)}({\bf\Delta}_{max}^2)+ \varepsilon{\cal
A}_1^{S(2)}({\bf\Delta}_{max}^2)+(1-\varepsilon){\cal A}_1^{V(2)}({\bf
\Delta}_{max}^2);\\
\label{35}
{\cal A}_1^{(1)}({\bf\Delta}_{max}^2)&=&N\biggl\{1+\frac{1}{
m_b}X_-+\frac{1}{m_b^2}\biggl(Y_-+M_\rho(M_\rho-m_q)
+\bar\Lambda(M_\rho+m_q)-\frac{1}{2}\tilde\Lambda\eta(5m_q+3M_\rho)
\nonumber\\
& & +\frac{1}{3}\left\langle\frac{{\bf
p}^2}{\bar\varepsilon_q+m_q} \right\rangle
\left(2\tilde\Lambda\eta-m_q-3M_\rho\right) \biggr)\biggr\};\\
\label{36} {\cal A}_1^{S(2)}({\bf\Delta}_{max}^2)
&=&N\biggl\{ \frac{2}{m_b}(M_\rho-2\langle\bar\varepsilon_q\rangle)+
\frac{1}{m_b^2}\biggl(
-2(M_\rho+\bar\Lambda+3m_q)(M_\rho-2\langle\bar\varepsilon_q\rangle)
\nonumber\\
& & -\frac{2}{3}\left\langle \frac{{\bf
p}^2}{\bar\varepsilon_q+m_q}\right\rangle(M_\rho R +m_q(1+3R)+
\bar\Lambda(1+R))
+\frac{2}{3}\langle{\bf p}^2\rangle(1+3R)\nonumber\\
& &-\tilde\Lambda \eta\left( M_\rho(2+R)
-\langle\bar\varepsilon_q\rangle(7+3R)
+\bar\Lambda(3+ R)  +\frac{1}{3}\left\langle\frac{{\bf p}^2}{\bar
\varepsilon_q}\right\rangle(1+R)\right)\biggr\};\\
\label{37}
{\cal A}_1^{V(2)}({\bf\Delta}_{max}^2) &=&N \biggl\{\frac{1}{m_b}
(4RZ_1+2RZ_2) +\frac{1}{m_b^2}\biggl( \frac{4}{3}(m_q-3M_\rho)RZ_3
-10\langle{\bf p}^2\rangle\left(1-\frac{2}{3}R\right)\nonumber\\
& & +\left\langle\frac{{\bf p}^2}{\bar\varepsilon_q+m_q}\right\rangle
\left( 2M_\rho
\left(2+\frac{13}{3}R\right) +8m_q\left(1-\frac{4}{3}R\right)
\right)\nonumber\\
& & +2\left(2m_q(1-R)-2M_\rho R
+\eta\frac{\beta_\rho^2}{\beta^2}\frac{\Delta\beta^2}{\beta}R\right)Z_.
\nonumber\\
& &+2\tilde\Lambda\eta\biggl(\left\langle\frac{1}{\bar\varepsilon_q
+m_q}\right\rangle\tilde\Lambda\eta\left(M_\rho\left(1-
\frac{5}{3}R\right)
-\bar\Lambda +m_q\left(1-\frac{10}{3}R\right)\right)\nonumber\\
& &+2(3\bar\Lambda -M_\rho-\langle\bar\varepsilon_q\rangle)
-\tilde\Lambda\eta\left(1 -\frac{10}{3}R\right)
-\frac{10}{3}\left\langle\frac{{\bf p}^2}
{\bar\varepsilon_q}\right\rangle\biggr)\biggr)\biggr\};
\end{eqnarray}
\begin{eqnarray}\label{38}{\cal A}_2({\bf\Delta}_{max}^2)&=&{\cal
A}_2^{(1)}({\bf\Delta}_{max}^2)+ \varepsilon{\cal
A}_2^{S(2)}({\bf\Delta}_{max}^2)+(1-\varepsilon){\cal A}_2^{V(2)}({\bf
\Delta}_{max}^2);\\
\label{39}
{\cal A}_2^{(1)}({\bf\Delta}_{max}^2)&=&N\biggl\{1+\frac{1}{
m_b}X_-+\frac{1}{m_b^2}\biggl(Y_--M_\rho(3M_\rho-m_q)
+\bar\Lambda(M_\rho-m_q)-\frac{1}{2}\tilde\Lambda\eta(M_\rho-7m_q)
\nonumber\\
& & -2\tilde\Lambda^2\eta^2+\frac{1}{3}\left\langle\frac{{\bf
p}^2}{\bar\varepsilon_q+m_q} \right\rangle
\left(2\tilde\Lambda\eta-m_q+5M_\rho\right) \biggr)\biggr\};\\
\label{40} {\cal A}_1^{S(2)}({\bf\Delta}_{max}^2)
&=& {\cal A}_1^{S(2)}({\bf\Delta}_{max}) +N\biggl\{
\frac{1}{m_b^2}4M_\rho(M_\rho-2\langle\bar\varepsilon_q
\rangle)\biggr\};\\
\label{41}
{\cal A}_2^{V(2)}({\bf\Delta}_{max}^2) &=&N \biggl\{\frac{1}{m_b}
(4RZ_1+2RZ_2) +\frac{1}{m_b^2}\biggl( \frac{4}{3}(M_\rho-m_q)RZ_3
-2\langle{\bf p}^2\rangle\left(1-\frac{10}{3}R\right)\nonumber\\
& & +\left\langle\frac{{\bf p}^2}{\bar\varepsilon_q+m_q}\right\rangle
\left(\frac{2}{3}M_\rho \left(1-11R\right) +\frac{2}{3}\bar\Lambda
+2m_q(1+R) \right)\nonumber\\ & &
+2\left(m_q(1+3R)-3M_\rho R
+\eta\frac{\beta_\rho^2}{\beta^2}\frac{\Delta\beta^2}{\beta}R\right)Z_.
\nonumber\\
& &+2\tilde\Lambda\eta\biggl(-\left\langle\frac{1}{\bar\varepsilon_q
+m_q}\right\rangle\tilde\Lambda\eta\left(M_\rho\left(1+
\frac{5}{3}R\right)
+\bar\Lambda +m_q\left(3+\frac{10}{3}R\right)\right)\nonumber\\
& &+3\bar\Lambda -M_\rho-\langle\bar\varepsilon_q\rangle
+\tilde\Lambda\eta\left(3 +\frac{10}{3}R\right)
-\frac{1}{3}\left\langle\frac{{\bf p}^2}
{\bar\varepsilon_q}\right\rangle\biggr)\biggr)\biggr\};
\end{eqnarray}
\begin{eqnarray}\label{42}{\cal V}({\bf\Delta}_{max}^2)&=&{\cal
V}^{(1)}({\bf\Delta}_{max}^2)+ \varepsilon{\cal
V}^{S(2)}({\bf\Delta}_{max}^2)+(1-\varepsilon){\cal V}^{V(2)}({\bf
\Delta}_{max}^2);\\
\label{43}
{\cal V}^{(1)}({\bf\Delta}_{max}^2)&=&N\biggl\{1+\frac{1}{
m_b}X_++\frac{1}{m_b^2}\biggl(Y_-+M_\rho(M_\rho-m_q) +\frac{1}{2}m_q^2
+\bar\Lambda(2m_q-M_\rho)+\frac{7}{2}\tilde\Lambda\eta M_\rho
\nonumber\\
& &-4\frac{1}{3}\langle{\bf p}^2\rangle
-\frac{1}{3}\left\langle\frac{{\bf p}^2}{\bar\varepsilon_q+m_q}
\right\rangle (M_\rho-m_q) \biggr)\biggr\};\\
\label{44} {\cal V}^{S(2)}({\bf\Delta}_{max}^2)
&=&N\biggl\{ -\frac{2}{m_b}(M_\rho-2\langle\bar\varepsilon_q\rangle)+
\frac{1}{m_b^2}\biggl(
-2(M_\rho-\bar\Lambda-m_q)(M_\rho-2\langle\bar\varepsilon_q\rangle)
\nonumber\\
& & +\frac{2}{3}\left\langle \frac{{\bf
p}^2}{\bar\varepsilon_q+m_q}\right\rangle(M_\rho(1-R) +m_q(2-3R)-
\bar\Lambda R)
+\frac{2}{3}\langle{\bf p}^2\rangle(2-3R)\nonumber\\
& &+\tilde\Lambda \eta\left(- M_\rho(1+2R)
+6\langle\bar\varepsilon_q\rangle R+
2\bar\Lambda(1- R)  -\frac{2}{3}\left\langle\frac{{\bf p}^2}{\bar
\varepsilon_q}\right\rangle(2-R)\right)\biggr\};\\
\label{45}
{\cal V}^{V(2)}({\bf\Delta}_{max}^2) &=&N \biggl\{\frac{1}{m_b}
(8RZ_1-2(1-R)Z_2) +\frac{1}{m_b^2}\biggl(-
\frac{8}{3}(M_\rho-m_q)RZ_3 +20\langle{\bf p}^2\rangle\nonumber\\
& & -\left\langle\frac{{\bf p}^2}{\bar\varepsilon_q+m_q}\right\rangle
\left( 2M_\rho \left(\frac{11}{3}-4R\right) +\frac{22}{3}\bar\Lambda
+4m_q\left(5+2R\right) \right)\nonumber\\ & &
+2\left(\bar\Lambda +\left(M_\rho R
+\eta\frac{\beta_\rho^2}{\beta^2}\frac{\Delta\beta^2}{\beta}\right)
(1-R)+m_q(1+R)\right)Z_2 \nonumber\\
& &+\tilde\Lambda^2\eta^2\biggl(17-
\left\langle\frac{1}{\bar\varepsilon_q
+m_q}\right\rangle\left(\frac{16}{3}M_\rho
+\frac{19}{3}\bar\Lambda +17m_q \right)
\biggr)\biggr)\bigg\},
\end{eqnarray}
where $N=\left(\frac{2\beta_B\beta_{\pi(\rho)}}{
 \beta_B^2+\beta_{\pi(\rho)}^2}\right)^{3/2}
 =\left(\frac{\beta_{\pi(\rho)}}{\beta_B}\eta\right)^{3/2}$ is due to
 the normalization of Gaussian wave functions in (\ref{14});
$\bar\varepsilon_q =\sqrt{{\bf p}^2+m_q^2+\tilde\Lambda^2\eta^2}
$, i.~e.  the energies of light quarks in final light meson
acquire additional contribution from the recoil momentum. The
averaging $\langle\dots\rangle$ is taken over the Gaussian wave
functions of $B$ and $\pi(\rho)$ mesons, so it can be carried out
analytically. For example,
\begin{equation}\label{q}\langle\bar\varepsilon_q\rangle= N^{-1}\int
\frac{d^3p}{(2\pi)^3}\bar\Psi_{\pi(\rho)}({\bf
p})\bar\varepsilon_q(p)\Psi_B({\bf p}) = \frac{1}{ \sqrt\pi}\frac{\bar
m_q^2}{\beta_{\pi(\rho)}\sqrt\eta}e^zK_1(z),
\end{equation}
where $\bar m_q^2=m_q^2+\tilde\Lambda^2\eta^2$ and $K_1(z)$ is the
modified Bessel function; $z={\bar m_q^2/(2\eta\beta_{\pi(\rho)}^2)}$.
Analogous expressions can be obtained for the other matrix elements in
(\ref{30})--(\ref{45}).

We have introduced the following notations:
\begin{eqnarray*}
X_1&=&\frac{2}{3}\left\langle\frac{{\bf p}^2}{\bar\varepsilon_q
+m_q}\right\rangle -\frac{1}{2}\tilde\Lambda\eta,\\
X_\pm&=&\frac{1}{3}\left\langle\frac{{\bf p}^2}{\bar\varepsilon_q
+m_q}\right\rangle +\frac{1}{2}\tilde\Lambda\eta \pm(M_\rho-m_q),\\
Y_\pm &=& -\frac{11}{24}\langle{\bf p}^2\rangle
+\frac{1}{2}(M_{\pi(\rho)}^2-m_q^2) \pm\frac{1}{2}\tilde\Lambda
\eta^2\left(\frac{1}{4}\tilde\Lambda+ \frac{\beta_{\pi(\rho)}^2}
{\beta^2}\frac{\Delta\beta^2}{\beta}\right),\\
Z_1& =& \frac{1}{3}\left\langle\frac{{\bf p}^2}{(\bar\varepsilon_q
+m_q)^2}\right\rangle(\bar\Lambda +M_{\pi(\rho)}+3m_q)
-\left\langle\frac{{\bf p}^2}{\bar\varepsilon_q +m_q}\right\rangle,\\
Z_2 &=& \tilde\Lambda\eta\left(\frac{1}{3}\left\langle \frac{{\bf
p}^2}{\bar\varepsilon_q(\bar\varepsilon_q+m_q)}\right\rangle
+\left\langle\frac{1}{\bar\varepsilon_q+m_q)}\right\rangle
(\bar\Lambda+M_{\pi(\rho)}+3m_q)-3\right),\\
Z_3 &=& \left\langle\frac{{\bf p}^2}{(\bar\varepsilon_q +m_q)^2}
\right\rangle(\bar\Lambda+M_{\pi(\rho)}+3m_q)
\end{eqnarray*}
and
$$R=\frac{m_b}{\varepsilon_b(\Delta_{max})+m_b}=\frac{1}{\sqrt{5}+2}.$$

\section{RESULTS AND DISCUSSION}

Using the parameters of Gaussian wave functions (\ref{14}) and the
value of the mixing coefficient of vector and scalar confining
potentials $\varepsilon=-1$ \cite{8} in the expressions (\ref{24}),
(\ref{30})--(\ref{33}) for the $B\to \pi$ transition form factor
$f_+(0)$ and the  eqs.~(\ref{25}), (\ref{26}), (\ref{34})--(\ref{45})
for the $B\to \rho$ transition form factors $A_1(0)$, $A_2(0)$ and
$V(0)$ we get
\begin{eqnarray}\label{46} f_+^{B\to \pi}(0)&=&0.20\pm 0.02\qquad
V^{B\to \rho}(0) =0.29\pm 0.03\nonumber\\ A_1^{B\to\rho}(0)&=&0.26\pm
0.03\qquad A_2^{B\to \rho}(0)=0.31\pm 0.03.\end{eqnarray}
The theoretical uncertainty in (\ref{46}) results mostly from the
approximation of the wave functions by Gaussians (\ref{14}) and does
not exceed 10\% of form factor values. In ref.~\cite{10a} we have
presented the results for $B\to\pi(\rho)e\nu$ decay form factors up to
the first order in $1/m_b$ expansion. The found values of form factors
\cite{10a} are very close to (\ref{46}), this indicates that the
second order correctons in $1/m_Q$ are small (less than 5\% of  form
factor values).

We compare our results (\ref{46}) for the form factors of $B\to
\pi(\rho)e\nu$ decays with the predictions of quark models
\cite{16,17}, QCD sum rules \cite{18,19,19a} and lattice calculations
\cite{l1,l2} in Table 1. There is an
agreement between our value of $f_+^{B\to \pi}(0)$ and QCD sum rule
and
lattice predictions. Our $B\to \rho e\nu$ form factors agree with
lattice
 and QCD sum rule ones \cite{19a}, while they are  approximately
1.5 times less than QCD sum rule results of refs.~\cite{18,19}.

To calculate the $B\to \pi(\rho)$ semileptonic decay rates it is
necessary to determine the $q^2$-dependence of the form factors.
Analysing the ${\bf\Delta}_{max}^2$ dependence of the expressions
(\ref{a1})--(\ref{a12}), (\ref{24})--(\ref{27}) for the form factors
$f_+$, $A_1$, $A_2$ and $V$, we find that the $q^2$-dependence of
these form factors near $q^2=0$ is given by

\begin{eqnarray} \label{47}
f_+(q^2)&=&\frac{M_B+M_\pi}{2\sqrt{M_B
M_\pi}}\tilde\xi(w){\cal F}_+({\bf\Delta}_{max}^2),\\
\label{48}
A_1(q^2)&=&\frac{2\sqrt{M_B M_\rho}}{M_B+M_\rho}\frac{1}{2}(1+w)
\tilde\xi(w){\cal A}_1 ({\bf\Delta}_{max}^2),\\
\label{49}
A_2(q^2)&=&\frac{M_B+M_\rho}{2\sqrt{M_B M_\rho}}\tilde\xi(w){\cal A}_2
({\bf\Delta}_{max}^2),\\
\label{50}
V(q^2)&=&\frac{M_B+M_\rho}{2\sqrt{M_B
M_\rho}}\tilde\xi(w){\cal V} ({\bf\Delta}_{max}^2),\end{eqnarray}
where $w=\frac{M_B^2+M_{\pi(\rho)}^2-q^2}{2M_B M_{\pi(\rho)}}$; ${\cal
F}_+({\bf\Delta}_{max}^2)$  and ${\cal A}_{1,2}({\bf
\Delta}_{max}^2)$, ${\cal V}({\bf\Delta}_{max}^2)$ are defined
by (\ref{24})--(\ref{26}), (\ref{30})--(\ref{45}). We have introduced
the function
\begin{equation} \label{51} \tilde\xi(w)=\left({2\over
w+1}\right)^{1/2}\exp
\left(-\eta\frac{\tilde\Lambda^2}{\beta^2_B}\frac{w-1}{w+1}\right),
\end{equation}
which in the limit of infinitely heavy  quarks in the
initial and final mesons coinsides with the Isgur-Wise function of our
model \cite{8}. In this limit eqs.~(\ref{47})--(\ref{50}) reproduce
the leading order prediction of HQET~\cite{9}.

It is important to note that the form factor $A_1$ in (\ref{48}) has a
different $q^2$-dependence than the other form factors (\ref{47}),
(\ref{49}), (\ref{50}). In the quark models it is usually assumed the
pole \cite{16} or exponential \cite{17} $q^2$-behaviour for all form
factors. However, the recent QCD sum rule analysis indicates that the
form factor $A_1$ has $q^2$-dependence different from other form
factors \cite{18,19,19a}. In \cite{19} it even decreases with the
increasing $q^2$ as
\begin{equation} \label{52} A_1(q^2)\simeq
\left(1-\frac{q^2}{M_b^2}\right)A_1(0) \simeq \frac{2M_B
M_\rho}{(M_B+M_\rho)^2}(1+w)A_1(0). \end{equation}
Such behaviour corresponds to replacing $\tilde\xi(w)$ in (\ref{48})
by $\tilde\xi(w_{max})$.

We have calculated the decay rates of $B\to \pi(\rho)e\nu$ using our
form factor values at $q^2=0$ and the $q^2$-dependence
(\ref{47})--(\ref{51}) in the whole kinematical region (model A). We
have also used the pole dependence for form factors $f_+(q^2)$,
$A_2(q^2)$, $V(q^2)$ and $A_1(q^2)=\frac{2M_B
M_\rho}{(M_B+M_\rho)^2}(1+w) \frac{A_1(0)}{ 1-q^2/m_P^2}$ (model B),
which corresponds to replacing the function $\tilde\xi(w)$ (\ref{51})
by the pole form factor. The results are presented in Table 2 in
comparison with the quark model \cite{16,17}, QCD sum rule
\cite{18,19} and lattice (for $B\to\pi e\nu$) \cite{l1} predictions.
Lattice accuracy, at present, is not enough to estimate $B\to\rho
e\nu$
rates \cite{l3}. We see that our results for the above
mentioned models A and B of form factor $q^2$-dependence coincide
within errors.  The ratio of the rates $\Gamma(B\to \rho
e\nu)/\Gamma(B\to \pi e\nu)$ is considerably reduced in our model
compared to the BSW \cite{16} and ISGW \cite{17} models with the
simple pole or exponential $q^2$-behaviour of all form factors.
Meanwhile our prediction for this ratio is in agreement with QCD sum
rule results \cite{18,19}. The absolute values of the rates
$\Gamma(B\to\pi e\nu)$ and $\Gamma(B\to\rho e\nu)$ in our model are
close to those from QCD sum rules \cite{19}. The predictions for the
rates with longitudinally and transversely polarized $\rho$ meson
differ considerably in these approaches. This is mainly due to
different $q^2$-behaviour of $A_1$ (see (\ref{48}), (\ref{52}) or
pole dominance model \cite{16}). Thus the measurement of the
ratios $\Gamma(B \to\rho e\nu)/\Gamma(B\to\pi e\nu)$ and
$\Gamma_L/\Gamma_T$ should provide the test of $q^2$-dependence of
$A_1$ and may discriminate between these approaches.

The differential decay spectra $\frac{1}{\Gamma}\frac{d\Gamma}{d x}$
for $B\to \pi(\rho)$ semileptonic transitions, where
$x=\frac{E_l}{M_B}$
and $E_l$ is the lepton energy, are presented in ref.~\cite{10a} (see
Fig.~3).

We can use our results for $V$ and $A_1$ to test the HQET relation
\cite{20} between the form factors of the semileptonic and rare
radiative decays of $B$ mesons. Isgur and Wise \cite{20} have shown
that in the limit of infinitely heavy $b$-quark mass an exact relation
connects the form factors $V$ and $A_1$ with the rare radiative decay
$B\to\rho\gamma$ form factor $F_1$ defined by:
\begin{eqnarray} \label{53}
\langle \rho(p_\rho,e)\vert \bar ui\sigma _{\mu \nu}q^\nu P_Rb\vert
B(p_B)\rangle&=&i\epsilon_{\mu \nu \tau \sigma }e^{*\nu }p_B^\tau
p_\rho^\sigma F_1(q^2)\nonumber\\ & &+\big[e_\mu
^*(M_B^2-M_\rho^2)-(e^* q)(p_B+p_\rho)_\mu\big]G_2(q^2).\end{eqnarray}
This relation is valid for $q^2$ values sufficiently close to
$q^2_{max}=(M_B-M_\rho)^2$ and reads:
\begin{equation} \label{54}
F_1(q^2)=\frac{q^2+M_B^2-M_\rho^2}{ 2M_B}
\frac{V(q^2)}{
M_B+M_\rho}+\frac{M_B+M_\rho}{2M_B}A_1(q^2).\end{equation}
It has been argued in \cite{21,22,19}, that in these processes the
soft contributions dominate over the hard perturbative ones, and thus
the Isgur-Wise relations (\ref{54}) could be extended to the whole
range of $q^2$. In \cite{10} we developed $1/m_b$ expansion for the
rare radiative decay form factor $F_1(0)$ using the same ideas as in
the present discussion of semileptonic decays. It was shown that
Isgur-Wise relation (\ref{54}) is satisfied in our model at leading
order of $1/m_b$ expansion. The found value of the form factor of rare
radiative decay $B\to\rho\gamma$ up to the second order in $1/m_b$
expansion is \cite{10}
\begin{equation} \label{55}
F_1^{B\to\rho}(0)=0.26\pm 0.03.\end{equation}
Using (\ref{54}) and the values of form factors (\ref{46}) we
find
\begin{equation} \label{56}
F_1^{B\to\rho}(0)=0.27\pm 0.03,\end{equation}
which is in accord with (\ref{55}). Thus we conclude that $1/m_b$ and
$1/m_b^2$ corrections do not break the Isgur-Wise relation (\ref{54})
in our model.

\section{CONCLUSIONS}

We have presented in detail the method of the calculation of
electroweak decay matrix elements for transitions between different
meson states, based on the quasipotential approach in quantum field
theory. It has been shown that the heavy-to-heavy decay matrix element
can be expanded in inverse powers of the initial and final heavy quark
masses at the point of zero recoil of the final meson. On the other
hand, the heavy-to-light decay matrix element can be expanded in
inverse powers of the  initial heavy quark mass and
large recoil momentum of final light meson at the point of maximum
recoil of final meson. As a result the expansion of the heavy-to-light
decay matrix element in inverse powers of initial heavy quark mass
arises. This method permits the calculation of various radiative
and weak decays of heavy mesons with the complete account of
relativistic effects.

This method has been applied to the investigation of  the
semileptonic decays of $B$ mesons into light mesons. The recoil
momentum of final $\pi(\rho)$ meson is large compared to the
$\pi(\rho)$ mass almost in the whole kinematical range. This requires
the completely relativistic treatment of these decays. On the other
hand, the presence of large recoil momentum, which for $q^2=0$ is of
order $m_b/2$, allows for the $1/m_b$ expansion of weak decay
matrix element at this point. The contributions to this expansion come
both from  the heavy $b$-quark mass and large recoil momentum of the
light final meson.

We have performed the $1/m_b$ expansion of
the semileptonic decay form factors at $q^2=0$ up to the second order.
The $q^2$-dependence of the form factors near $q^2=0$ has been
determined. It has been found that the axial form factor $A_1$ has a
$q^2$-behaviour different from other form factors (see
(\ref{47})--(\ref{50})).  This is in agreement with recent QCD sum
rule results \cite{18,19,19a}. The ratios $\Gamma(B\to\rho
e\nu)/\Gamma(B\to\pi e\nu)$ and $\Gamma_L/\Gamma_T$ are very sensitive
to the $q^2$-dependence of $A_1$, and thus their experimental
measurement may discriminate between different approaches.

We have considered the relation between semileptonic decay form
factors and the rare radiative decay form factor \cite{20}, obtained
in the limit of the infinitely heavy $b$-quark. It has been found that
in our model $1/m_b$ corrections do not violate this relation.

\bigskip
\noindent {\bf ACKNOWLEDGEMENTS}
\nopagebreak

\medskip

\noindent We express our gratitude to B.~A.~Arbuzov, M.~A.~Ivanov,
J.~G.~K\"orner, V.~A.~Matveev, M.~Neubert, V.~I.~Savrin, B.~Stech, A.
Vainshtein for the interest in our work and helpful discussions of the
results. This research was supported in part by the Russian Foundation
for Fundamental Research under Grant No.94-02-03300-a and by the
Interregional Centre for Advanced Studies.

\begin{appendix}
\section{ APPENDIX: HEAVY-TO-LIGHT
SEMILEPTONIC DECAY FORM FACTORS AT $q^2=0$ POINT}

\noindent a) $B\to\pi e\nu$ decay form factor
\nopagebreak
\begin{eqnarray}\label{a1} f_+^{(1)}(0)&=&\sqrt{\frac{E_\pi}{ M_B}}
\int\frac{d^3p}{ (2\pi)^3} \bar\Psi_\pi\left({\bf
p}+\frac{2\varepsilon_q}{ E_\pi+M_\pi}{\bf\Delta}_{max}\right)
\sqrt{\frac{\varepsilon_q(p+\Delta_{max})+m_q}{
2\varepsilon_q(p+\Delta_{max})}} \nonumber \\
& &\times \sqrt{\frac{\varepsilon_b(p)+m_b}{2m_b}}
\biggl\{1+\frac{M_B-E_\pi}{\varepsilon_q(p+\Delta_{max})+m_q}
+\frac{({\bf p\Delta}_{max})}{{\bf\Delta}_{max}^2}
\biggl(\frac{\varepsilon_q(p +\Delta_{max})-m_q}{2m_b}\nonumber\\
& & + (M_B-E_\pi)\left(\frac{1}{\varepsilon_q(p +\Delta_{max})+m_q}+
\frac{1}{2m_b}\right)\biggr)
+(p_x^2+p_y^2)\Biggl(\frac{E_\pi-M_\pi}{2m_b(\varepsilon_q(p+
\Delta_{max}) +m_q)}\nonumber \\ & &\times \left(
\frac{1}{\varepsilon_q(p)+m_q}
-\frac{1}{\varepsilon_q(p+\Delta_{max})+m_q}\right)
+\frac{M_B-E_\pi}{E_\pi+M_\pi}\left(\frac{1}{\varepsilon_q(p
+\Delta_{max}) +m_q} -\frac{1}{2m_b}\right)\nonumber\\ & &\times
\left( \frac{1}{\varepsilon_q(p)+m_q}
-\frac{1}{\varepsilon_q(p+\Delta_{max}) +m_q}\right)\biggr)\Biggr\}
\Psi_B({\bf p}),\\
\label{a2}
 f_+^{S(2)}(0)&=&\sqrt{\frac{E_\pi}{ M_B}} \int\frac{d^3p}{
(2\pi)^3} \bar\Psi_\pi\left({\bf p}+\frac{2\varepsilon_q}{
E_\pi+M_\pi}{\bf\Delta}_{max}\right)
\sqrt{\frac{\varepsilon_q(p+\Delta_{max})+m_q}{
2\varepsilon_q(p+\Delta_{max})}} \nonumber \\
& &\times \Biggl\{\left(\frac{\varepsilon_q(\Delta_{max})-m_q}{
\varepsilon_q(\Delta_{max})+m_q}-
\frac{M_B-E_\pi}{\varepsilon_q(\Delta_{max})+m_q}\right)
\frac{1}{\varepsilon_q(
\Delta_{max})}\left(M_\pi-2\varepsilon_q\left(p+
\frac{2\varepsilon_q}{E_\pi+M_\pi}\Delta_{max}\right)\right)\nonumber\\
& & +\frac{({\bf p\Delta}_{max})}{2{\bf \Delta}_{max}^2}\biggl[\biggl(
\frac{\varepsilon_q(\Delta_{max})-m_q}{\varepsilon_q(\Delta_{max})
(\varepsilon_q(\Delta_{max})+m_q)} \nonumber \\ & & -
(M_B-E_\pi)\left(\frac{1}{
\varepsilon_q(\Delta_{max})(\varepsilon_q(\Delta_{max})+m_q)}+
\frac{1}{m_b(\varepsilon_b(\Delta_{max})+m_b)}\right)\biggr)
\nonumber \\
& & \times\left(M_B+M_\pi-\varepsilon_b(p)-
\varepsilon_q(p)-2\varepsilon_q\left(p+ \frac{2\varepsilon_q}{
E_\pi+M_\pi}\Delta_{max}\right)\right)\nonumber \\
& & - \frac{\varepsilon_q(\Delta_{max})-m_q}{2m_b(\varepsilon_b(
\Delta_{max})+m_b)}(M_B-\varepsilon_b(p)-\varepsilon_q(p))
+\frac{M_B-E_\pi}{2m_b\varepsilon_q(\Delta_{max})}\nonumber \\ & &
\times\frac{\varepsilon_q(\Delta_{max})-
m_q}{\varepsilon_q(\Delta_{max})
+m_q}
\left(M_\pi-2\varepsilon_q\left(p+\frac{2\varepsilon_q}{E_\pi
+M_\pi}\Delta_{max}\right)\right)\biggr]\nonumber\\
& & +\frac{p_x^2+p_y^2}{2(\varepsilon_q(p) +m_q)}\biggl[\biggl(
\frac{E_\pi-M_\pi}{\varepsilon_q(\Delta_{max})(\varepsilon_q(p+
\Delta_{max})+m_q)^2} \nonumber\\
& & -\frac{M_B-E_\pi}{E_\pi+M_\pi}\biggl(
\frac{1}{\varepsilon_q(\Delta_{max})(\varepsilon_q(\Delta_{max})+m_q)}
- \frac{1}{m_b(\varepsilon_b(\Delta_{max})+m_b)}\biggr)\biggr)
\nonumber\\
& &\times\left(M_B+M_\pi-\varepsilon_b(p)-
\varepsilon_q(p)-2\varepsilon_q\left(p+ \frac{2\varepsilon_q}{
E_\pi+M_\pi}\Delta_{max}\right)\right)\nonumber\\
& & +\frac{(E_\pi-M_\pi)(M_B-\varepsilon_b(p)-\varepsilon_q(p))}{
m_b(\varepsilon_b(\Delta_{max})+m_b)(\varepsilon_q(\Delta_{max})
+m_q)}- \frac{M_B-E_\pi}{(E_\pi+M_\pi)m_b\varepsilon_q(\Delta_{max})}
\nonumber\\
& &\times\frac{\varepsilon_q(\Delta_{max})-m_q}
{\varepsilon_q(\Delta_{max})+m_q}\left(M_\pi- 2\varepsilon_q\left(p+
\frac{2\varepsilon_q}{
E_\pi+M_\pi}\Delta_{max}\right)\right)\biggr]\Biggr\} \Psi_B({\bf
p}),\\
\label{a3}
f_+^{V(2)}(0)&=&\sqrt{\frac{E_\pi}{ M_B}} \int\frac{d^3p}{
(2\pi)^3} \bar\Psi_\pi\left({\bf p}+\frac{2\varepsilon_q}{
E_\pi+M_\pi}{\bf\Delta}_{max}\right)
\sqrt{\frac{\varepsilon_q(p+\Delta_{max})+m_q}{
2\varepsilon_q(p+\Delta_{max})}} \nonumber \\
& & \times \Biggr\{-\frac{p_x^2+p_y^2}{(\varepsilon_q(p)
+m_q)}\left(\frac{1}{\varepsilon_q(p)+m_q}
-\frac{1}{\varepsilon_q(\Delta_{max})+m_q}\right)\nonumber\\
& & \times \biggl(\frac{E_\pi -M_\pi}{\varepsilon_q(p+
\Delta_{max})+m_q} \biggl(\frac{1}{\varepsilon_b(\Delta_{max}) +m_b}
+\frac{1}{\varepsilon_q(\Delta_{max})+m_q}\biggr)\nonumber \\
& &-\frac{M_B-E_\pi}{E_\pi+M_\pi}\biggl(
\frac{1}{\varepsilon_q(\Delta_{max})+m_q} -
\frac{1}{\varepsilon_b(\Delta_{max})+m_b}\biggr)\biggr) \nonumber\\
& &\times\left(M_B+M_\pi-\varepsilon_b(p)-
\varepsilon_q(p)-2\varepsilon_q\left(p+ \frac{2\varepsilon_q}{
E_\pi+M_\pi}\Delta_{max}\right)\right)\nonumber\\
& & +\frac{({\bf p\Delta}_{max})}{{\bf \Delta}_{max}^2}\frac{1}{
\varepsilon_q(p)+m_q} \biggl[ (\varepsilon_q(\Delta_{max}) -m_q)
\nonumber\\ & & \times
\left(\frac{1}{\varepsilon_b(\Delta_{max})+m_b}- \frac{m_q}
{\varepsilon_q(\Delta_{max}) (\varepsilon_q(\Delta_{max})+m_q)}
\right)\nonumber \\
& & + (M_B-E_\pi)\left(\frac{1}{ \varepsilon_q(\Delta_{max})+m_q}
+\frac{1}{\varepsilon_b(\Delta_{max})+m_b}\right)\biggr]
\nonumber \\
& & \times\left(M_B+M_\pi-\varepsilon_b(p)-
\varepsilon_q(p)-2\varepsilon_q\left(p+ \frac{2\varepsilon_q}{
E_\pi+M_\pi}\Delta_{max}\right)\right)\nonumber \\
& & +\frac{{\bf p}^2}{\varepsilon_q(p)+m_q}\biggl[
\frac{M_B-M_\pi-\varepsilon_b(p)-
\varepsilon_q(p)+2\varepsilon_q\left(p+ \frac{2\varepsilon_q}{
E_\pi+M_\pi}\Delta_{max}\right)}{2\varepsilon_q(\Delta_{max})
(\varepsilon_q(\Delta_{max})+m_q)}  \nonumber \\
& & -\frac{M_B-E_\pi}{\varepsilon_q(\Delta_{max})+m_q}
\frac{M_B+M_\pi-\varepsilon_b(p)-
\varepsilon_q(p)-2\varepsilon_q\left(p+ \frac{2\varepsilon_q}{
E_\pi+M_\pi}\Delta_{max}\right)}{2\varepsilon_q(\Delta_{max})
(\varepsilon_q(\Delta_{max})+m_q)}  \nonumber \\
& & +\left(\frac{1}{\varepsilon_q(\Delta_{max})+m_q} -\frac{1}{2m_b}
\left(1-\frac{M_B-E_\pi}{\varepsilon_q(\Delta_{max})+m_q}\right)
\right)\frac{M_B-\varepsilon_b(p)-\varepsilon_q(p)}{\varepsilon_b(
\Delta_{max})+m_b} \biggr]\Biggr\}\Psi_B({\bf p}),
\end{eqnarray}

\noindent b) $B\to \rho e\nu$ decay form factors
\nopagebreak
\begin{eqnarray}\label{a4} A_1^{(1)}(0)&=&\frac{2\sqrt{M_BM_\rho}}{
M_B+M_\rho} \sqrt{\frac{E_\rho}{ M_\rho}} \int\frac{d^3p}{ (2\pi)^3}
\bar\Psi_\rho\left({\bf p}+\frac{2\varepsilon_q}{
E_\rho+M_\rho}{\bf\Delta}_{max}\right)
\sqrt{\frac{\varepsilon_q(p+\Delta_{max})+m_q}{
2\varepsilon_q(p+\Delta_{max})}} \nonumber \\
& &\times \sqrt{\frac{\varepsilon_b(p)+m_b}{2m_b}}
\Biggl\{1+\frac{1}{2m_b(\varepsilon_q(p+\Delta_{max})+m_q)}
\biggl((E_\rho-M_\rho)\frac{p_x^2+p_y^2}{\varepsilon_q(p)+m_q}
\nonumber\\
& & -\frac{1}{3}{\bf p}^2 -({\bf p\Delta}_{max})\biggr)\Biggr\}
\Psi_B({\bf p}),\\
\label{a5} A_1^{S(2)}(0)&=&\frac{2\sqrt{M_BM_\rho}}{
M_B+M_\rho} \sqrt{\frac{E_\rho}{ M_\rho}} \int\frac{d^3p}{ (2\pi)^3}
\bar\Psi_\rho\left({\bf p}+\frac{2\varepsilon_q}{
E_\rho+M_\rho}{\bf\Delta}_{max}\right)
\sqrt{\frac{\varepsilon_q(p+\Delta_{max})+m_q}{
2\varepsilon_q(p+\Delta_{max})}} \nonumber \\
& &\times \Biggl\{\frac{\varepsilon_q(\Delta_{max})-m_q}{
\varepsilon_q(\Delta_{max})+m_q}\frac{1}{\varepsilon_q(\Delta_{max})}
\left(M_\rho
-2\varepsilon_q\left(p+\frac{2\varepsilon_q}{E_\rho+M_\rho}\Delta_{max}
\right)\right) \nonumber\\ & &
-\frac{(p_x^2+p_y^2)(E_\rho-M_\rho)}{2m_b(\varepsilon_q(p)+m_q)
(\varepsilon_q(\Delta_{max})+m_q)}
\biggl[\frac{1}{\varepsilon_b(\Delta_{max})+m_b}\nonumber\\
& & \times\left(M_B+M_\rho-\varepsilon_b(p)-
\varepsilon_q(p)-2\varepsilon_q\left(p+ \frac{2\varepsilon_q}{
E_\rho+M_\rho}\Delta_{max}\right)\right)\nonumber\\
& & +\frac{1}{\varepsilon_q(\Delta_{max})}(M_B-\varepsilon_b(p)
-\varepsilon_q(p))\biggr] + \frac{({\bf p\Delta}_{max})}
{{\bf\Delta}_{max}^2} \frac{\varepsilon_q(\Delta_{max})-m_q}{2}
\biggl[\biggl( \frac{1}{m_b(\varepsilon_b(\Delta_{max})+m_b)}
\nonumber\\ & &
+\frac{1}{\varepsilon_q(\Delta_{max})(\varepsilon_q(\Delta_{max})
+m_q)}\biggr) \left(M_B+M_\rho-\varepsilon_b(p)-
\varepsilon_q(p)-2\varepsilon_q\left(p+ \frac{2\varepsilon_q}{
E_\rho+M_\rho}\Delta_{max}\right)\right)\nonumber\\
& &+\frac{1}{m_b\varepsilon_q(\Delta_{max})}(M_B-\varepsilon_b(p)-
\varepsilon_q(p))\biggr]\Biggr\}  \Psi_B({\bf p}),\\
\label{a6} A_1^{V(2)}(0)&=&\frac{2\sqrt{M_BM_\rho}}{
M_B+M_\rho} \sqrt{\frac{E_\rho}{ M_\rho}} \int\frac{d^3p}{ (2\pi)^3}
\bar\Psi_\rho\left({\bf p}+\frac{2\varepsilon_q}{
E_\rho+M_\rho}{\bf\Delta}_{max}\right)
\sqrt{\frac{\varepsilon_q(p+\Delta_{max})+m_q}{
2\varepsilon_q(p+\Delta_{max})}} \nonumber \\
& &\times \Biggl\{\frac{(p_x^2+p_y^2)(E_\rho-M_\rho)}
{(\varepsilon_q(p)+m_q)(\varepsilon_q(\Delta_{max})+m_q)}
\biggl(\frac{1}{(\varepsilon_q(p)+m_q)(\varepsilon_b(\Delta_{max})
+m_b)} \nonumber\\
& &+\frac{1}{(\varepsilon_q(\Delta_{max})+m_q)^2}\biggr)
\left(M_B+M_\rho-\varepsilon_b(p)-
\varepsilon_q(p)-2\varepsilon_q\left(p+ \frac{2\varepsilon_q}{
E_\rho+M_\rho}\Delta_{max}\right)\right)\nonumber\\ & &
+\frac{{\bf p}^2}{\varepsilon_q(p)+m_q}\biggl[ \frac{1}{2\varepsilon_q
(\Delta_{max})(\varepsilon_q(\Delta_{max})+m_q)} \biggl(
\frac{\varepsilon_q(\Delta_{max})-
m_q}{\varepsilon_q(\Delta_{max})+m_q}
\nonumber\\
& &\times\biggl(M_\rho
 -2\varepsilon_q\left(p+\frac{2\varepsilon_q}{E_\rho+
M_\rho}\Delta_{max} \right)\biggr)
 -(M_B-\varepsilon_b(p)-\varepsilon_q(p))\biggr)\nonumber\\
& &-\frac{1}
{\varepsilon_b(\Delta_{max})+m_b}\left(\frac{1}{3(\varepsilon_q(
\Delta_{max})+m_q)}+\frac{1}{m_b}\right)\nonumber\\
& &\times\left(M_\rho-
2\varepsilon\left(p+\frac{2\varepsilon_q}{E_\rho+
M_\rho}\Delta_{max} \right)\right)\biggr]- \frac{({\bf
p\Delta}_{max})} {{\bf\Delta}_{max}^2}
(\varepsilon_q(\Delta_{max})-m_q) \nonumber \\
& &\times\biggl[\frac{1}{2\varepsilon_q(\Delta_{max})(\varepsilon_q
(\Delta_{max})+m_q)}\biggl(2(M_B-\varepsilon_b(p)-\varepsilon_q(p))
\nonumber\\
& &+\frac{\varepsilon_q(\Delta_{max})-m_q}{\varepsilon_q(\Delta_{max})
+m_q}\left(M_B+M_\rho-\varepsilon_b(p)-
\varepsilon_q(p)-2\varepsilon_q\left(p+ \frac{2\varepsilon_q}{
E_\rho+M_\rho}\Delta_{max}\right)\right)\biggr)\nonumber\\
& & +\frac{1}{\varepsilon_q(p)+m_q}\left(
\frac{1}{\varepsilon_b(\Delta_{max})+m_b}+
\frac{m_q}{\varepsilon_q(\Delta_{max})(\varepsilon_q(\Delta_{max})
+m_q)}\right) \nonumber\\
& & \times \left(M_B+M_\rho-\varepsilon_b(p)-
\varepsilon_q(p)-2\varepsilon_q\left(p+ \frac{2\varepsilon_q}{
E_\rho+M_\rho}\Delta_{max}\right)\right)\biggr]\Biggr\} \Psi_B({\bf
p}),
\end{eqnarray}
\begin{eqnarray}
\label{a7} A_2^{(1)}(0)&=&\frac{M_B+M_\rho}{2\sqrt{M_BM_\rho}}
\frac{2\sqrt{E_\rho M_\rho}}{E_\rho+ M_\rho} \int\frac{d^3p}{
(2\pi)^3} \bar\Psi_\rho\left({\bf p}+\frac{2\varepsilon_q}{
E_\rho+M_\rho}{\bf\Delta}_{max}\right)
\sqrt{\frac{\varepsilon_q(p+\Delta_{max})+m_q}{
2\varepsilon_q(p+\Delta_{max})}} \nonumber \\
& & \times \sqrt{\frac{\varepsilon_b(p)+m_b}{2m_b}}
\Biggl\{1+\frac{M_\rho}{M_B}\left(1-\frac{E_\rho+M_\rho}
{\varepsilon_q(p+\Delta_{max})+m_q}\right) -\frac{p_z^2}{2m_b
(\varepsilon_q(p+\Delta_{max})+m_q)}\nonumber\\
& & +\frac{p_x^2+p_y^2}{(\varepsilon_q(p+\Delta_{max})+m_q)
(\varepsilon_q(p)+m_q)}\left(\frac{E_\rho+M_\rho}{2m_b}
+\frac{M_\rho}{M_B}\right) \nonumber\\
& & -\frac{({\bf p\Delta}_{max})}{{\bf\Delta}_{max}^2}\left(
\frac{\varepsilon_q(p+\Delta_{max})-m_q}{2m_b} +\frac{M_\rho}{M_B}
\left(1+\frac{E_\rho}{m_b}\right)\right)\Biggr\} \Psi_B({\bf p}),\\
\label{a8} A_2^{S(2)}(0)&=&\frac{M_B+M_\rho}{2\sqrt{M_BM_\rho}}
\frac{2\sqrt{E_\rho M_\rho}}{E_\rho+ M_\rho} \int\frac{d^3p}{
(2\pi)^3} \bar\Psi_\rho\left({\bf p}+\frac{2\varepsilon_q}{
E_\rho+M_\rho}{\bf\Delta}_{max}\right)
\sqrt{\frac{\varepsilon_q(p+\Delta_{max})+m_q}{
2\varepsilon_q(p+\Delta_{max})}} \nonumber \\
& &\times \Biggl\{\left(\frac{\varepsilon_q(\Delta_{max})-m_q}{
\varepsilon_q(\Delta_{max})+m_q}+2\frac{M_\rho}{M_B}\right)
\frac{1}{\varepsilon_q(\Delta_{max})} \left(M_\rho
-2\varepsilon_q\left(p+\frac{2\varepsilon_q}{E_\rho+M_\rho}\Delta_{max}
\right)\right) \nonumber\\
& & -\frac{p_x^2+p_y^2}{2m_b(\varepsilon_q(p)+m_q)}
\biggl[\frac{1}{\varepsilon_b(\Delta_{max})+m_b}\nonumber\\ & &
\times\left(M_B+M_\rho-\varepsilon_b(p)-
\varepsilon_q(p)-2\varepsilon_q\left(p+ \frac{2\varepsilon_q}{
E_\rho+M_\rho}\Delta_{max}\right)\right)\nonumber\\
& & +\frac{1}{\varepsilon_q(\Delta_{max})}(M_B-\varepsilon_b(p)
-\varepsilon_q(p))\biggr] + \frac{({\bf p\Delta}_{max})}
{{\bf\Delta}_{max}^2} \frac{1}{2}\biggl[\biggl(\frac{
\varepsilon_q(\Delta_{max})}{m_b(\varepsilon_b(\Delta_{max})+m_b)}
\nonumber\\
& &+\frac{1}{\varepsilon_q(\Delta_{max})+m_q}\biggr)
\left(M_B+M_\rho-\varepsilon_b(p)-
\varepsilon_q(p)-2\varepsilon_q\left(p+ \frac{2\varepsilon_q}{
E_\rho+M_\rho}\Delta_{max}\right)\right)\nonumber\\
& &+\frac{1}{m_b}(M_B-\varepsilon_b(p)-
\varepsilon_q(p))\biggr]\Biggr\}  \Psi_B({\bf p}),\\
\label{a9} A_2^{V(2)}(0)&=&\frac{M_B+M_\rho}{2\sqrt{M_BM_\rho}}
\frac{2\sqrt{E_\rho M_\rho}}{E_\rho+ M_\rho} \int\frac{d^3p}{
(2\pi)^3} \bar\Psi_\rho\left({\bf p}+\frac{2\varepsilon_q}{
E_\rho+M_\rho}{\bf\Delta}_{max}\right)
\sqrt{\frac{\varepsilon_q(p+\Delta_{max})+m_q}{
2\varepsilon_q(p+\Delta_{max})}} \nonumber \\
& &\times \Biggl\{\frac{p_x^2+p_y^2}
{(\varepsilon_q(p)+m_q)(\varepsilon_q(\Delta_{max})+m_q)}
\biggl(\frac{E_\rho+M_\rho}{(\varepsilon_q(p)+m_q)(\varepsilon_b(
\Delta_{max}) +m_b)} \nonumber\\ &
&+\frac{1}{\varepsilon_q(\Delta_{max})+m_q}\biggr)
\left(M_B+M_\rho-\varepsilon_b(p)-
\varepsilon_q(p)-2\varepsilon_q\left(p+ \frac{2\varepsilon_q}{
E_\rho+M_\rho}\Delta_{max}\right)\right)\nonumber\\
& & -\frac{{\bf p}^2}{\varepsilon_q(p)+m_q}\biggl[
\frac{1}{2\varepsilon_q
(\Delta_{max})(\varepsilon_q(\Delta_{max})+m_q)} \nonumber\\
& &\times\biggl(M_B+M_\rho-\varepsilon_b(p)-\varepsilon_q(p)
-2\varepsilon_q\left(p+ \frac{2\varepsilon_q}{E_\rho+
 M_\rho}\Delta_{max} \right)\biggr) \nonumber\\
& &+\frac{1}{\varepsilon_b(\Delta_{max})+m_b}
\left(\frac{1}{3(\varepsilon_q(\Delta_{max})+m_q)}
+\frac{1}{m_b}\right)\left(M_\rho-2\varepsilon\left(p
+\frac{2\varepsilon_q}{E_\rho+ M_\rho}\Delta_{max}
\right)\right)\biggr]\nonumber\\
& &- \frac{({\bf p\Delta}_{max})}{{\bf\Delta}_{max}^2}
\biggl[\frac{1}{\varepsilon_q(p)+m_q} \left(
\frac{\varepsilon_q(\Delta_{max})+m_q}{\varepsilon_b(\Delta_{max})
+m_b}+\frac{m_q}{2\varepsilon_q(\Delta_{max})} \right)\nonumber\\
& &\times \left(M_B+M_\rho-\varepsilon_b(p)-
\varepsilon_q(p)-2\varepsilon_q\left(p+ \frac{2\varepsilon_q}{
E_\rho+M_\rho}\Delta_{max}\right)\right)\biggr)\nonumber\\
& & +\frac{1}{\varepsilon_q(\Delta_{max})}\biggl(3(M_B-
\varepsilon_b(p)
-\varepsilon_q(p))\nonumber\\
& &-\left(M_\rho-2\varepsilon_q\left(p+\frac{2\varepsilon_q}{
E_\rho+M_\rho}\Delta_{max}\right)\right)\biggr)\biggr]\Biggr\}
\Psi_B({\bf
p}),\end{eqnarray}
\begin{eqnarray}
\label{a10} V^{(1)}(0)&=&\frac{M_B+M_\rho}{2\sqrt{M_BM_\rho}}
\int\frac{d^3p}{(2\pi)^3} \bar\Psi_\rho\left({\bf p}
+\frac{2\varepsilon_q}{ E_\rho+M_\rho}{\bf\Delta}_{max}\right)
\frac{2\sqrt{E_\rho M_\rho}}{\varepsilon_q(p+\Delta_{max})+m_q}
\sqrt{\frac{\varepsilon_q(p+\Delta_{max})+m_q}{
2\varepsilon_q(p+\Delta_{max})}} \nonumber \\
& & \times \sqrt{\frac{\varepsilon_b(p)+m_b}{2m_b}}
\Biggl\{1+\frac{p_x^2+p_y^2}{E_\rho+M_\rho}\left(
\frac{\varepsilon_q(p+\Delta_{max})+m_q}{2m_b(\varepsilon_q(p)+m_q)}
-\frac{1}{\varepsilon_q(p+\Delta_{max})+m_q}\right)\nonumber\\
& & +\frac{({\bf p\Delta}_{max})}{{\bf\Delta}_{max}^2}\left( 1-
\frac{\varepsilon_q(p+\Delta_{max})+m_q}{2m_b} \right)
\Biggr\} \Psi_B({\bf p}),\\
\label{a11} V^{S(2)}(0)&=&\frac{M_B+M_\rho}{2\sqrt{M_BM_\rho}}
\int\frac{d^3p}{(2\pi)^3} \bar\Psi_\rho\left({\bf p}
+\frac{2\varepsilon_q}{ E_\rho+M_\rho}{\bf\Delta}_{max}\right)
\frac{2\sqrt{E_\rho M_\rho}}{\varepsilon_q(p+\Delta_{max})+m_q}
\sqrt{\frac{\varepsilon_q(p+\Delta_{max})+m_q}{
2\varepsilon_q(p+\Delta_{max})}} \nonumber \\
& &\times \Biggl\{-\frac{1}{\varepsilon_q(\Delta_{max})}
\left(M_\rho -2\varepsilon_q\left(p+\frac{2\varepsilon_q}
{E_\rho+M_\rho}\Delta_{max} \right)\right)
+\frac{p_x^2+p_y^2}{2m_b(E_\rho+M_\rho)(\varepsilon_q(p)+m_q)}
\nonumber\\
& & \times\biggl[\frac{\varepsilon_q(p+\Delta_{max})-m_q}
{\varepsilon_q(\Delta_{max})}
\left(M_\rho -2\varepsilon_q\left(p+\frac{2\varepsilon_q}
{E_\rho+M_\rho}\Delta_{max} \right)\right) \nonumber\\
& & -\frac{\varepsilon_q(\Delta_{max})+m_q}{\varepsilon_b
(\Delta_{max}) +m_b} \left(M_B+M_\rho-\varepsilon_b(p)-
\varepsilon_q(p)-2\varepsilon_q\left(p+ \frac{2\varepsilon_q}{
E_\rho+M_\rho}\Delta_{max}\right)\right)\biggr]\nonumber\\ & & -
\frac{({\bf p\Delta}_{max})} {{\bf\Delta}_{max}^2}
\biggl[\frac{1}{2}\left(\frac{1}{\varepsilon_q(\Delta_{max})} -
\frac{\varepsilon_q(\Delta_{max})+m_q}{ m_b
(\varepsilon_b(\Delta_{max}) +m_b)}\right)\nonumber\\ & &\times
\left(M_B+M_\rho-\varepsilon_b(p)-
\varepsilon_q(p)-2\varepsilon_q\left(p+ \frac{2\varepsilon_q}{
E_\rho+M_\rho}\Delta_{max}\right)\right)\nonumber\\ &
&-\frac{\varepsilon_q(\Delta_{max})-m_q}{2m_b
\varepsilon_q(\Delta_{max})} \left(M_\rho
-2\varepsilon_q\left(p+\frac{2\varepsilon_q}
{E_\rho+M_\rho}\Delta_{max} \right)\right) \biggr]\Biggr\}
\Psi_B({\bf p}),\\ \label{a12}
V^{V(2)}(0)&=&\frac{M_B+M_\rho}{2\sqrt{M_BM_\rho}}
\int\frac{d^3p}{(2\pi)^3} \bar\Psi_\rho\left({\bf p}
+\frac{2\varepsilon_q}{E_\rho+M_\rho}{\bf\Delta}_{max}\right)
\frac{2\sqrt{E_\rho M_\rho}}{\varepsilon_q(p+\Delta_{max})+m_q}
\sqrt{\frac{\varepsilon_q(p+\Delta_{max})+m_q}{
2\varepsilon_q(p+\Delta_{max})}} \nonumber \\ & &\times
\Biggl\{\frac{2(p_x^2+p_y^2)}{(E_\rho+M_\rho)
(\varepsilon_q(p)+m_q)}\biggl(\frac{\varepsilon_q(\Delta_{max})+m_q}
{(\varepsilon_q(p)+m_q)(\varepsilon_b(\Delta_{max})+m_b)} \nonumber\\
& &-\frac{1}{\varepsilon_q(\Delta_{max})+m_q}\biggr)
\left(M_B+M_\rho-\varepsilon_b(p)-
\varepsilon_q(p)-2\varepsilon_q\left(p+ \frac{2\varepsilon_q}{
E_\rho+M_\rho}\Delta_{max}\right)\right)\nonumber\\ & & -\frac{{\bf
p}^2}{2m_b(\varepsilon_q(p)+m_q)}\biggl[
\frac{1}{\varepsilon_b(\Delta_{max})+m_b}(M_B-\varepsilon_b(p)
-\varepsilon_q(p))\nonumber\\ & &
+\frac{1}{\varepsilon_q(\Delta_{max})+m_q}
\left(M_B+M_\rho-\varepsilon_b(p)-\varepsilon_q(p)-
2\varepsilon\left(p+\frac{2\varepsilon_q}{E_\rho+ M_\rho}\Delta_{max}
\right)\right)\biggr] \nonumber\\ & & + \frac{({\bf p\Delta}_{max})}
{{\bf\Delta}_{max}^2} \frac{1}{\varepsilon_q(p)+m_q}\left(1-
\frac{\varepsilon_q(\Delta_{max})+m_q}{\varepsilon_b
(\Delta_{max})+m_b}\right)\nonumber\\
& &\times\left(M_B+M_\rho-\varepsilon_b(p)-\varepsilon_q(p)-
2\varepsilon\left(p+\frac{2\varepsilon_q}{E_\rho+ M_\rho}\Delta_{max}
\right)\right)\Biggr\} \Psi_B({\bf p}),
\end{eqnarray}
where the superscripts ``(1)" and ``(2)" correspond to Figs.~1 and 2,
$S$ and $V$ --- to the scalar and vector potentials of $q\bar
q$-interaction;
\begin{equation}\label{a13}\vert {\bf\Delta}_{max}\vert=\frac{M_B^2-
M_{\pi(\rho)}^2}{ 2M_B};\qquad E_{\pi(\rho)}=\sqrt{M_{\pi(\rho)}^2+
{\bf\Delta}_{max}^2}= \frac{M_B^2+ M_{\pi(\rho)}^2}{ 2M_B};
\end{equation}
and $z$-axis is chosen in the direction of ${\bf\Delta}_{max}$.

\end{appendix}

\nonfrenchspacing

\newpage
\noindent {\bf TABLE~1} Semileptonic $B\to \pi$ and $B\to \rho$ decay
form factors.

\bigskip
\begin{tabular}{ccccc}
\hline
Ref. & $f_+^{B\to\pi}(0)$ & $A_1^{B\to\rho}(0)$ & $A_2^{B\to\rho}(0)$
& $V^{B\to\rho}(0)$ \\
\hline
our results & $0.20\pm 0.02$ & $0.26\pm 0.03$ & $0.31\pm 0.03$ &
$0.29\pm 0.03$ \\
\cite{16}$^a$ & 0.33 & 0.28 & 0.28 & 0.33 \\
\cite{17}$^a$ & 0.09 & 0.05 & 0.02 & 0.27 \\
\cite{18}$^b$ & $0.26\pm 0.02$ & $0.5\pm 0.1$ & $0.4\pm 0.2$ & $0.6\pm
0.2$ \\
\cite{19}$^b$ & $0.23\pm 0.02$ & $0.38\pm 0.04$ & $0.45\pm 0.05$ &
$0.45\pm 0.05$ \\
\cite{19a}$^b$ &   &$0.24\pm0.04$ &  & $0.28\pm 0.06$\\
\cite{l1}$^c$ & $0.35\pm0.08$ & $0.24\pm 0.12$ & $0.27\pm 0.80$ &
$0.53\pm 0.31$\\
\cite{l2}$^c$ & $0.30\pm 0.14\pm 0.05$ & $0.22\pm 0.05$ & $0.49\pm
0.21
\pm 0.05 $ & $0.37\pm 0.11$\\
\hline
\end{tabular}

\medskip
$^a$ quark models

$^b$ QCD sum rules

$^c$ lattice

\bigskip

\medskip

\bigskip

\noindent {\bf TABLE~2} Semileptonic decay rates $\Gamma(B\to\pi
e\nu)$, $\Gamma(B\to \rho e\nu)$  ($\times \vert
V_{ub}\vert^2\times 10^{12}$s${}^{-1}$) and the ratio of the rates for
longitudinally and transversely polarized $\rho$ meson.

\bigskip

\begin{tabular}{cccc}
\hline
Ref. & $\Gamma (B\to\pi e\nu)$ & $\Gamma(B\to\rho e\nu)$ &
$\Gamma_L/\Gamma_T$ \\
\hline
our results &   &   &   \\
model A & $3.0\pm 0.6$ & $5.4\pm 1.2$ & $0.5\pm 0.3$ \\
model B & $2.9\pm 0.6$ & $5.0\pm 1.2$ & $0.5\pm 0.3$ \\
\cite{16}$^a$ & 7.4 & 26 & 1.34 \\
\cite{17}$^a$ & 2.1 & 8.3 & 0.75 \\
\cite{18}$^b$ & $5.1\pm 1.1$ & $12\pm 4$ & $0.06\pm 0.02$ \\
\cite{19}$^b$ & $3.6\pm 0.6$ & $5.1\pm 1.0$ & $0.13\pm 0.08$ \\
\cite{l1}$^c$ & $8\pm 4$ &  &  \\
\hline
\end{tabular}

\medskip
$^a$ quark models

$^b$ QCD sum rules

$^c$ lattice

\bigskip

\bigskip

\bigskip

\noindent {\large\bf FIGURE CAPTIONS}

\smallskip
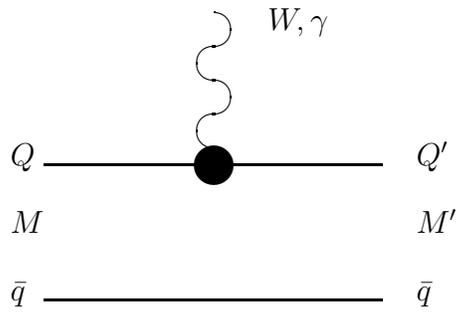
\noindent {\bf FIGURE~1} Lowest order vertex function

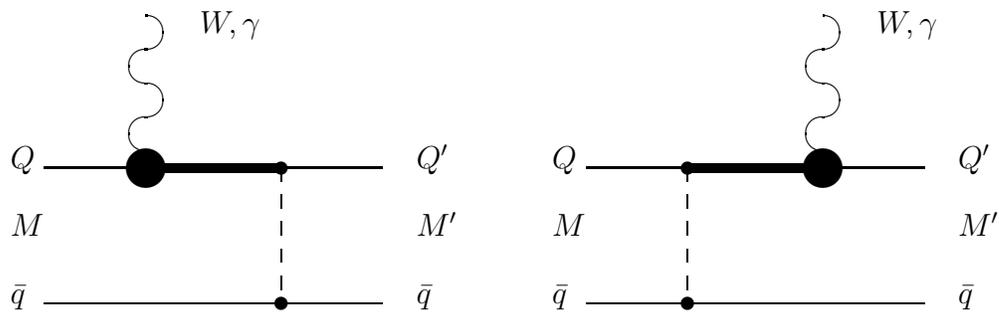
\noindent {\bf FIGURE~2} Vertex function with the account of the quark
interaction. Dashed line corresponds to the effective potential
(\ref{5}).  Bold line denotes the negative-energy part of the quark
propagator.

\vfill
\newpage
\unitlength=0.9mm
\large
\begin{picture}(150,150)
\put(10,100){\line(1,0){50}}
\put(10,120){\line(1,0){50}}
\put(35,120){\circle*{8}}
\multiput(32.5,130)(0,-10){2}{\begin{picture}(5,10)
\put(2.5,10){\oval(5,5)[r]}
\put(2.5,5){\oval(5,5)[l]}\end{picture}}
\put(5,120){$Q$}
\put(5,100){$\bar q$}
\put(5,110){$M$}
\put(65,120){$Q'$}
\put(65,100){$\bar q$}
\put(65,110){$M'$}
\put(43,140){$W,\gamma$}
\put(30,85){\Large\bf Fig. 1}
\put(10,20){\line(1,0){50}}
\put(10,40){\line(1,0){50}}
\put(25,40){\circle*{8}}
\put(25,40){\thicklines \line(1,0){20}}
\multiput(25,40.5)(0,-0.1){10}{\thicklines \line(1,0){20}}
\put(25,39.5){\thicklines \line(1,0){20}}
\put(45,40){\circle*{2}}
\put(45,20){\circle*{2}}
\multiput(45,40)(0,-4){5}{\line(0,-1){2}}
\multiput(22.5,50)(0,-10){2}{\begin{picture}(5,10)
\put(2.5,10){\oval(5,5)[r]}
\put(2.5,5){\oval(5,5)[l]}\end{picture}}
\put(5,40){$Q$}
\put(5,20){$\bar q$}
\put(5,30){$M$}
\put(65,40){$Q'$}
\put(65,20){$\bar q$}
\put(65,30){$M'$}
\put(33,60){$W,\gamma$}
\put(90,20){\line(1,0){50}}
\put(90,40){\line(1,0){50}}
\put(125,40){\circle*{8}}
\put(105,40){\thicklines \line(1,0){20}}
\multiput(105,40.5)(0,-0.1){10}{\thicklines \line(1,0){20}}
\put(105,39,5){\thicklines \line(1,0){20}}
\put(105,40){\circle*{2}}
\put(105,20){\circle*{2}}
\multiput(105,40)(0,-4){5}{\line(0,-1){2}}
\multiput(122.5,50)(0,-10){2}{\begin{picture}(5,10)
\put(2.5,10){\oval(5,5)[r]}
\put(2.5,5){\oval(5,5)[l]}\end{picture}}
\put(85,40){$Q$}
\put(85,20){$\bar q$}
\put(85,30){$M$}
\put(145,40){$Q'$}
\put(145,20){$\bar q$}
\put(145,30){$M'$}
\put(133,60){$W,\gamma$}
\put(70,5){\Large \bf Fig. 2}

\end{picture}

\end{document}